\documentclass[aps,prl,footinbib,floats,twocolumn,showpacs,superscriptaddress,longbibliography]{revtex4-1}

\usepackage{graphicx}
\usepackage{siunitx}
\usepackage{multirow}
\usepackage[normalem]{ulem}
\usepackage{color}

\usepackage{url}

\usepackage{breakurl}
\usepackage[breaklinks]{hyperref}

\begin{document}

\title{Socio-economic constraints to maximum human lifespan}

\author{Albert Sol\'e-Ribalta}
\affiliation{Internet Interdisciplinary Institute (IN3), Universitat Oberta de Catalunya, 08035 Barcelona}
\author{Javier Borge-Holthoefer}
\affiliation{Internet Interdisciplinary Institute (IN3), Universitat Oberta de Catalunya, 08035 Barcelona}

\begin{abstract}
The analysis of the demographic transition of the past century and a half, using both empirical data and mathematical models, has rendered a wealth of well-established facts, including the dramatic increases in life expectancy. Despite these insights, such analyses have also occasionally triggered debates which spill over many disciplines, from genetics, to biology, or demography. Perhaps the hottest discussion is happening around the question of maximum human lifespan, which --besides its fascinating historical and philosophical interest-- poses urgent pragmatic warnings on a number of issues in public and private decision-making. In this paper, we add to the controversy some results which, based on purely statistical grounds, suggest that the maximum human lifespan is not fixed, or has not reached yet a plateau. Quite the contrary, analysis on reliable data for over 150 years in more than 20 industrialized countries point at a sustained increase in the maximum age at death. Furthermore, were this trend to continue, a limitless lifespan could be achieved by 2102. Finally, we quantify the dependence of increases in the maximum lifespan on socio-economic factors. Our analysis indicates that in some countries the observed rising patterns can only be sustained by progressively larger increases in GDP, setting the problem of longevity in a context of diminishing returns.
\end{abstract}

\maketitle


\section{Background}


With an increasing reliability, abundance and availability of demographic data over the past century, scientists have managed to deliver and agree on a number of facts. First and foremost is the evidence for a sharp increase in life expectancy all along the 20th century \cite{wilmoth2000increase,oeppen2002broken}. With all the implications it bears --from social planning to climate change--, average human lifespan is not particularly revealing in a biological sense: certainly, the increases we have witnessed have more to do with our capacity to manipulate the environment (e.g. medical advances, a stable and balanced diet, labor conditions \cite{wilkinson1992income,falagas2009economic, wolfson1999relation}), than changes in the biology of the species. In this sense, scientific advances seem to have outrun evolution. However, as we live longer, a natural question arises: is there an inherent limit to human life? If so, are we approaching it?

These simple questions have triggered a long and hot debate, which is not settled. Part of the debate happens in the biological arena, where some theories on ageing processes conclude that these are inescapable \cite{kirkwood2000we} and ultimately fatal, e.g. antagonistic pleiotropy or disposable soma theory. On the contrary, others claim that there is no physical restriction preventing an indefinite postponement to senescence \cite{kirkwood2000we}. Indeed, ageing processes do not seem to be pervasive in nature \cite{garcia2015long,klapper1998longevity,salo2006power}. Lifespan may not be an absolute hard limit, but a function of other parameters, which include the environment where we live (in the broader sense) and our capacity to alter it \cite{kirkwood1977evolution,koubova2003does,kirkwood2005understanding}. While the evolution of life expectancy seems to support precisely this view --lifespan as a function of the environment--, at least on statistical grounds, we still lack a conclusive biological argument around maximum human lifespan.

In the meantime, scholars have turned to data again --this time to tackle the question of human life's upper-bounds, rather than expectation. In a recent work \cite{dong2016evidence}, it was claimed that human lifespan limit has already been reached, and such limit was set at 115. However, serious methodological concerns were quickly raised \cite{brown2017contesting,hughes2017many,lenart2017questionable,rozing2017there}, projecting a shadow of doubt and jeopardizing the main conclusions. We should keep in sight that estimates for lifespan limits --either in terms of life expectancy or maximum lifespan-- have been broken in the past, while others have not yet been exceeded \cite{maxLiveSpan120_2014}. From a statistically less naive point of view, implied distributions of raw maximum ages at death could be predicted directly in the context of parametric models for hazard functions: for Gompertz, Gamma-Gompertz, Makeham, Gamma-Gompertz-Makeham, or others \cite{finch1996maximum}, the distribution of maximum age within a birth cohort of, say, size $n$ at age 50 is readily available. However, the distinctive behaviour of distributions at their extreme tails have long been one of the reasons that most demographers reject Gompertz models as a description of survival to extreme ages: maximum lifespan is an extreme value statistic, and therefore particularly affected by sample size \cite{kirkwood1997origins}.

In this paper, we take a Big Data approach to analyze demographic and mortality data under the light of extreme value statistics. We do so for datasets of more than 20 countries, with over a century of reliable information. Results tell us that maximum lifespan is not fixed, but has rather evolved in a steep linear increasing tendency for the last 60 years --in correspondence with post-World War II general progress. Projecting the trends into the future, maximum lifespan might become unbounded within the next 100 years. These trends, we hypothesize, may have gone previously unnoticed due to the Simpson's paradox. In the second part of the article, we relate maximum lifespan extension to standardized Gross Domestic Product (GDP). Results evidence that, in some countries, the increment of the GDP has a diminishing returns effect on maximum lifespan. However, interestingly, the effect does not seem to be neither heavily related to the maximum lifespan nor to the absolute GDP value.

\section{Data Description}

We combine the information from the Human Mortality Database (HMD) \cite{mortalitydb}, the International Database on Longevity (IDL) \cite{maier2010supercentenarians} (1750-2014), and the Gerontology Wiki (GW) \cite{gerontology} for a representative set of countries. The HMD reports the number of deaths per calendar year and age up to 109 years, aggregating counts henceforth. This dataset does not specify the exact death date of individuals. The IDL and GW complement the HMD dataset. IDL reports data only for super-centenarians ($\ge 110$ years old) with the exact demise date. With similar aims, GW mostly overlaps with IDL, but comprehends a number of countries which are not in IDL.

The previous data are segmented in death cohorts, i.e. people who died on the same calendar year. Death cohorts guarantee that life conditions during the last period of life are shared among the population --opposite to birth cohorts, which are more suited in other settings. Additionally, death cohorts allow us to extend the analysis to year 2014. Death cohorts have been previously used in \cite{wilmoth2000increase,dong2016evidence,gbari2017extreme}.

On the merged dataset, we apply a block maxima method to each calendar year --death cohort--. Each block corresponds to a day. Since we do not have exact date of birth and death for individuals below 110 years, we randomly cast HMD data to a daily distribution, given the seasonal birth and death distribution obtained from UNdata (http://data.un.org/). On top of the random sample, we add the values for which death date is known ($\ge$ 110, IDL and GW). With this distribution at hand, the obtained samples are fitted, using maximum likelihood, to estimate the GEV distribution parameters. The reported parameters of the GEV distribution for each cohort are averaged over 200 random realizations to ensure statistical significant results.


\section{Results}

\begin{figure*}[bt!]
  \begin{center}
	 \begin{tabular}{c}
		\includegraphics[width=2\columnwidth]{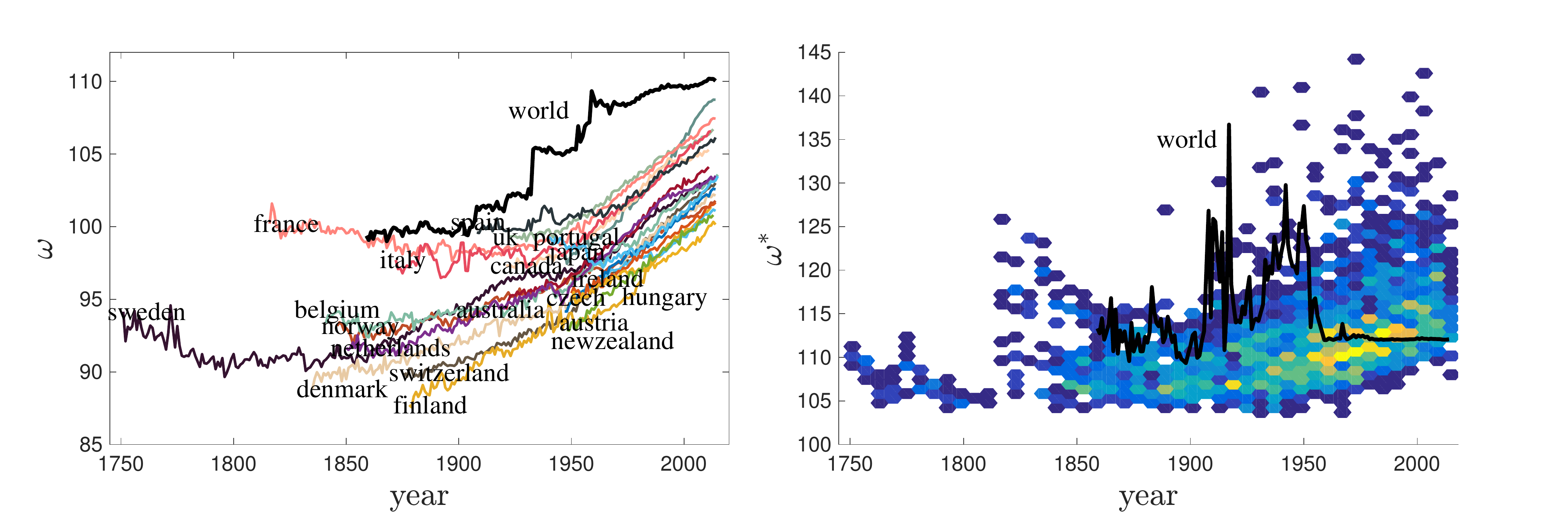}\\
		\includegraphics[width=2\columnwidth]{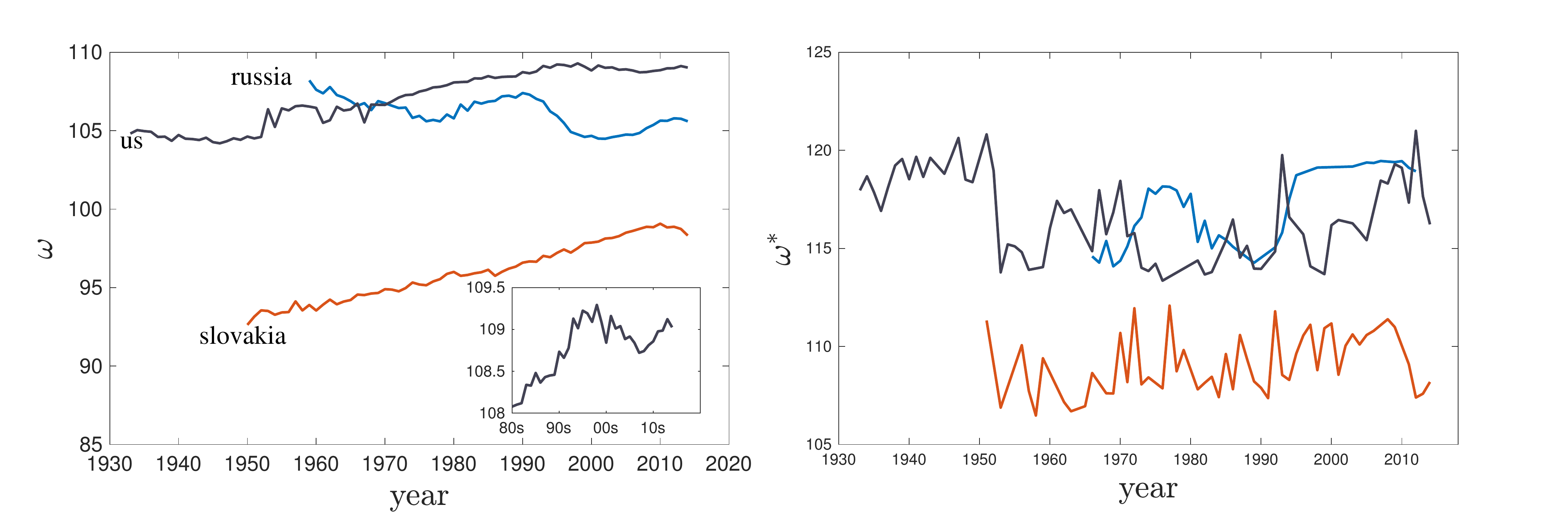}
 	 \end{tabular}
	\end{center}
	\caption{{\bf Expected and theoretical maximum lifespan.} Panel A and C report on the expected maximum lifespan for each cohort for all countries under scrutiny. Panel C shows the 3 countries where we observe an anomalous behavior: US, Russia and Slovakia. The expected maximum lifespan $\omega$ corresponds to the first moment of the GEV distribution that best fits empirical mortality data, including centenarians and super-centenarians. Notably, all trends show an growing tendency for maximum lifespan, with a clear linear regime from mid-20th century onwards. For the same period of time, panels B and D displays the theoretical upper-bound $\omega^{*}$ of the fitted GEV distribution that, in all cases, correspond to a reversed Weibull (see Methods). Considering the noise observed in the evolution of $\omega^{*}$ and the amount of countries analysed we present this evolution in terms of a heat-map. The intensity of the points correspond to the amount of countries that display this $\omega^{*}$ at that year. This theoretical extreme also presents a rising pattern. The evolution of $\omega^{*}$ for the anomalous countries is presented in panel D.}
	\label{fig1}
\end{figure*}

We start by analysing the existence of an inherent limit to longevity throughout the results presented in Figure~\ref{fig1}. Left panels report the expected maximum lifespan $\omega$, for the set of studied countries, as a function of calendar year.  Such estimate $\omega$ corresponds here to the first moment of fitted Generalized Extreme Value (GEV) distribution, see Methods for details. With few exceptions, presented in the bottom panels, maximum lifespan has monotonically risen for over 100 years. Furthermore, this increasing trend has consistently grown at a linear pace in the last 60 years, with correlation coefficients in the range $0.96 \leq r^{2} \leq 0.99$ for all countries. Despite the different slopes, which are indicative of the varied historical trajectories of the countries, the observed pattern is substantial and persistent. This linear behavior is perfectly compatible with the one reported for life expectancy \cite{oeppen2002broken} and mortality at advanced ages \cite{kannisto1994reductions}. Also, it is aligned with the increase in the variation in lifespan among survivors to older ages (e.g. 65 and above) \cite{engelman2014lifespan}.

Besides remarkably high Pearson coefficients for the period 1960-2014, the slopes of these linear trends belong to the narrow range $0.11 \pm 0.05$, implying a similar growing trend for these subset of countries. The difference in the intercept of such trend (vertical shifts) may be explained by the richness of the sample in the extreme values: smaller countries present less instances of extreme age than those with larger populations. Figure S1 of the Supplementary Information illustrates this point, confronting $\omega_t$ ($t \in [1960, 2014]$) with population at year $t$.

Countries with longer historical data show a smooth change of tendency from decreasing to increasing after 1850-1900. Without a deep analysis of the socio-economic and political situation of each country, the most plausible explanation for this convex shape is improved hygiene, public health, medicine, nutrition, and technology.
In any case, the sharply increasing trends obtained under the light of GEV statistics, after 1960, differ largely from the evolution of life expectancy, which is valuable to anticipate general trends (see Figure~S2 of the Supplementary Information) --but not extreme scores, which demand a different treatment.

The right panels of Figure~\ref{fig1} report on the theoretical estimation of maximum human lifespan \cite{gbari2017extreme,weon2009theoretical}, according to the fitted distribution. In particular, when the shape parameter of the fitted GEV distribution is lower than zero, $\xi < 0$, the distribution is equivalent to a reversed Weibull, characterized by a finite upper bound $\omega^{*}$; any non-negative value leads to a Gumbel ($\xi = 0$) or a Fr\'echet ($\xi > 0$) distribution, which lack a defined upper bound. The maximum likelihood estimates of the GEV parameters consistently indicates that the best distribution fitting our data is, unsurprisingly, the reversed Weibull and consequently the GEV has a well defined upper bound. Despite its noisy nature --slight changes in the shape parameter $\xi$ translate into large shifts of $\omega^{*}$--, this maximalist estimation shows as well an increasing behavior, with values ranging from 105 to more than 140. 

Additionally, top panels show the results considering all the available data for all countries. Note the behavior of $\omega$ and $\omega^{*}$ after the 60's. The expected maximum lifespan evidences a deterioration, and the theoretical upper bound shows a stationary behavior. None of these are observed in the per-country analyses. This discrepancy is a consequence of the Simpson's paradox, which can explain misleading results when data are considered at the aggregate level; it might also underlie the polemical results in \cite{dong2016evidence}.

As shown in Figure~\ref{fig1}, 3 out of the 23 countries show a deterioration of their maximum lifespan.

Noteworthy, US sets a global maximum $\omega = 109$ in year 1990 showing a decline afterwards despite not setting the theoretical global maximum (larger $\omega^{*}$). Inset of Figure~\ref{fig1} explores this particular empirical trend, zooming on the calendar years where the maximum is reached. As said, the expected maximum lifespan shows a non-monotonic behavior between the range $[1980, 2014]$, implying that the average value in the fitted GEV distribution appears to have stalled around age 109. This is also reflected in the distribution's upper bound $\omega^{*}$, which shows an oscillating trend.

The evolution of the maximum lifespan of Russia is clearly different from US. It shows an oscillating behavior that seem to follow the different socio-political and economical situation of the Soviet Union. It is worth letting know to the reader that before the Soviet Union dissolution only citizens, centenarians and super-centenarians that live in the future Russian territory are considered. The evolution we observe is as follows. A monotonic decrease in the 60's and 70's corresponding to the de-Stalinization era. Around the Stagnation era (1973-1985) we observe a monotonic increase of the lifespan. The Stagnation era was economically adverse but seem to provide enough stability to sustain elders life and their health. The maximum lifespan was increased by approximately two years. Afterwards a significant decrease, contemporary to the dissolution of the Soviet Union, is observed. This decrease reduced 5 years of maximum lifespan in 10 years. Intriguingly, this oscillation behavior is not observed in the life expectancy on age tables for the elder population. However, it is indeed reflected in the life expectancy at younger ages, the range $[0, 40]$. See Figure~1 of the Supplementary Material.

Slovakia also shows a decreasing trend after 2010. The reduction is not acute and may possibly be a delayed consequence of the economical crisis of 2007.

The evolution of the maximum lifespan of US, Slovakia, and the particular case of Russia, evidence to what extent the socio-economic and political situation affects life conditions of the elder and consequently affecting the observed maximum lifespan similarly to the control of the environment does to life expectancy \cite{caspari2006human}.


\begin{figure*}[bt!]
  \begin{center}
	 \begin{tabular}{c}
		\includegraphics[width=2\columnwidth]{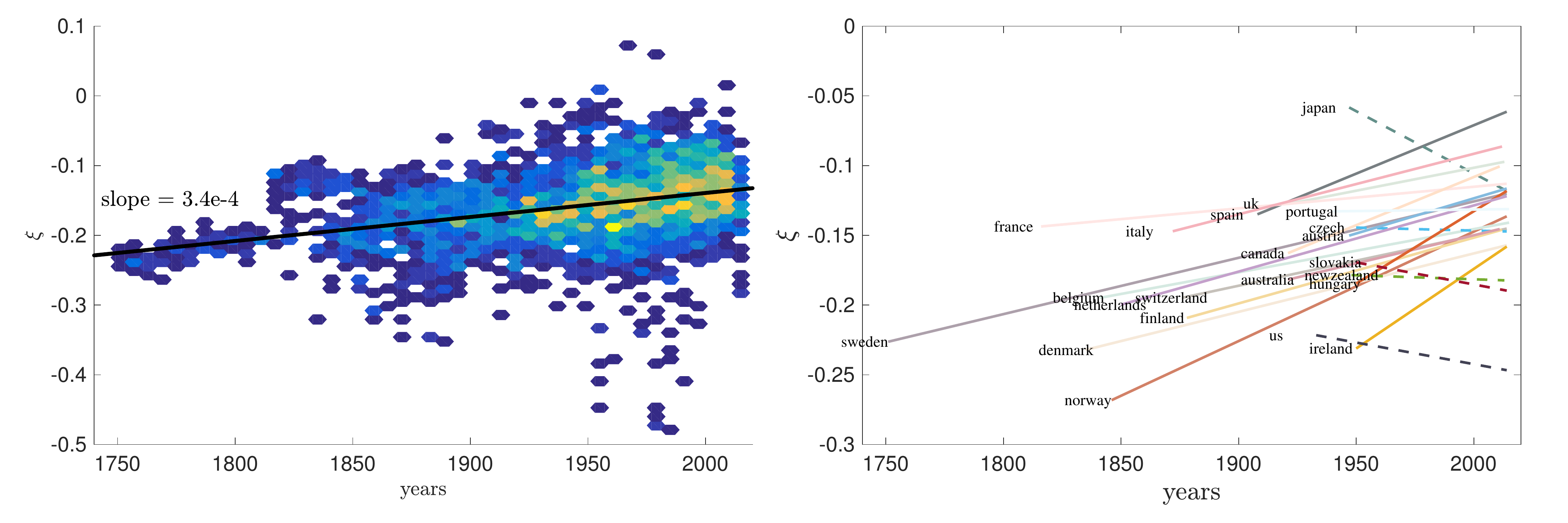}
 	 \end{tabular}
	\end{center}
	\caption{{\bf Temporal evolution of the shape parameter $\xi$.} Panel A show the evolution in form of a heat-map of the $\xi$ parameter of the fitted GEV distribution. The intensity of each point correspond to the amount of countries that display this $\xi$ parameter at that year. The solid black line corresponds to the linear fit of the evolution of the $\xi$ value. Panel B show the individual evolution of $\xi$ for each country in terms of its linear fit. The transparency of each line in panel B is proportional to the magnitude of the slope. Decreasing trends are displayed as dashed lines.}
	\label{fig:shape_param_linFit}
\end{figure*}

The generalized increasing trends of $\omega$ and $\omega^{*}$ suggest that, if there is, a natural limit to human lifespan has not been reached yet. To to further investigate the evolution of $\omega^{*}$, we resort on the shape parameter $\xi$ which dictates the theoretical upper bound of the GEV distribution. As can be observed in Figure~\ref{fig:shape_param_linFit}A, $\xi$ is strictly negative for all countries and so the upper bound (i.e. maximum lifespan) is well defined. The anecdotal four points outside the range, $\xi>0$, are regarded as noise since do not show neither any regularity nor tendency. The linear fit over the plot is a visual guide to indicate the general parsimonious increasing tendency of $\xi$ towards positive values.

Figure~\ref{fig:shape_param_linFit}b and Table~\ref{table:shape_evo} display, for each country, the linear fit of the evolution of $\xi$ values from the initial to the final calendar years available. If $\xi$ evolution presented a flat shape (slope of $\xi = 0$) and below 0 value, it would indicate an approximation of $\omega^{*}$ towards an asymptotic barrier or ``wall of death'' \cite{wachter2013age}. Quite the contrary, we observe a robust growing trend of the $\xi$ parameter towards positive values, with the notable exceptions of 5 countries. Nonetheless, the only significantly negative slope happens to be in Japan. We remark that, upon reaching $\xi \ge 0$, the distribution (now a Fr\'echet or Gumbel) ceases to be upper-bounded, which translates into a tendency for human lifespan to diverge. It seems then that, for most countries, $\xi$ shows a rising pattern --though admittedly at different pace. 

\begin{figure}[bt!]
  \begin{center}
	 \begin{tabular}{c}
		\includegraphics[width=1\columnwidth]{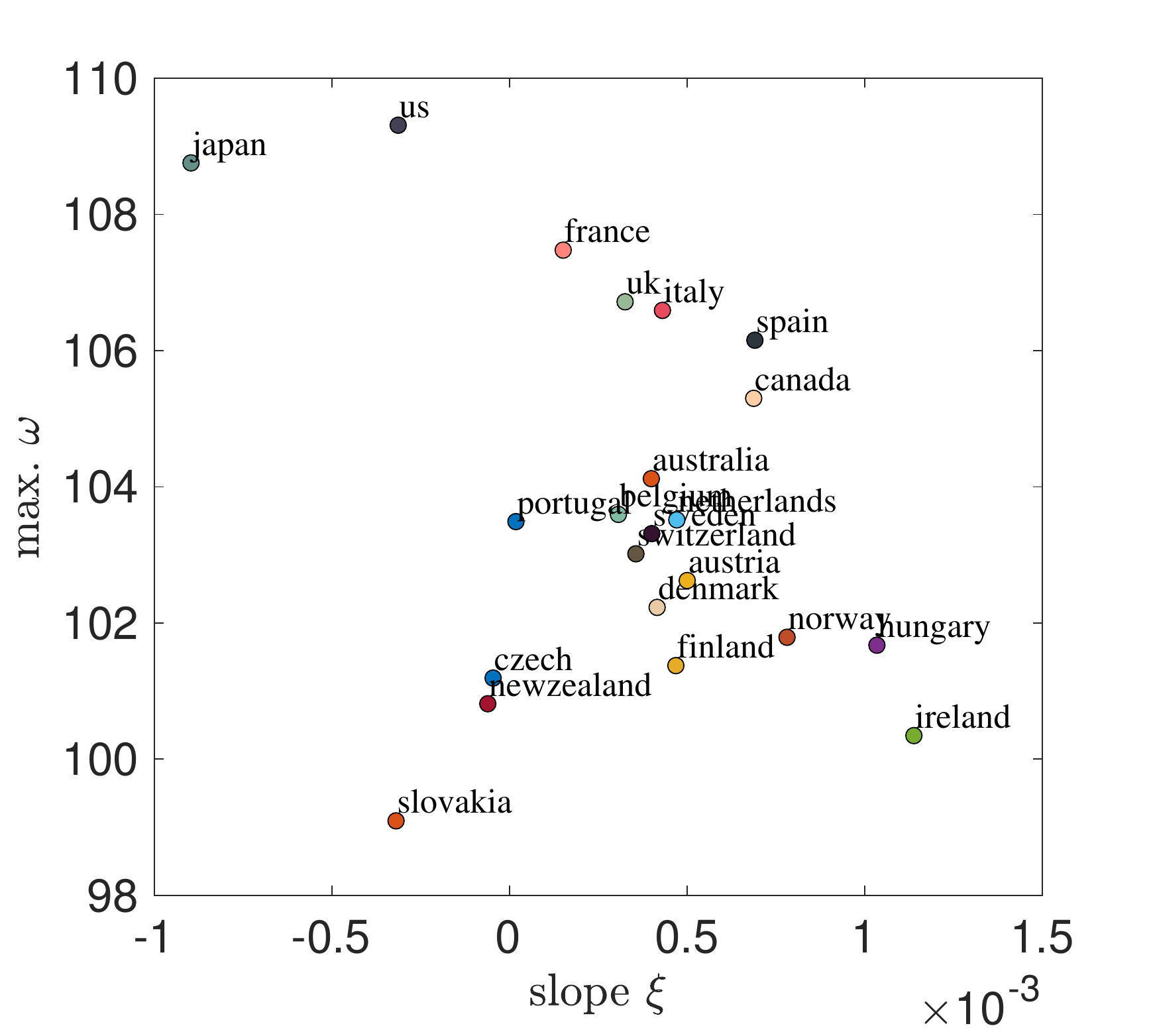}
 	 \end{tabular}
	\end{center}
	\caption{{\bf Relationship between expected maximum lifespan and evolution of $\xi$ parameter.} The maximum lifespan corresponds to the expected maximum obtained for all analysed years for each country. The slope of $\xi$ corresponds to the slope of linear fits in Figure~\ref{fig:shape_param_linFit} which are also reported in Table~\ref{table:shape_evo}.}
	\label{fig:relation_shape_maxw}
\end{figure}

A deeper analysis of Figure~\ref{fig:shape_param_linFit} and Table~\ref{table:shape_evo} show that two of the countries with larger lifespan present negative slopes. This might evidence a saturation and a subsequent decrease of the maximum lifespan once their maximum value is reached. To address this question, we present in Figure~\ref{fig:relation_shape_maxw} a scatter plot showing the relationship between the maximum obtained lifespan, $\mbox{max.}~\omega$, and the slope of the $\xi$ parameter. The overall obtained results indicate that there is no clear relationship between the reached maximum lifespan the observed slope in the $\xi$ parameter. Thus the observed decrease {seems to remove:(must)} depend on other factors rather than the current maximum lifespan.

Finally, we use forecasting techniques to uncover long-term trends (rather than {\it predictions}). At this point, the central question is not about how long humans can live, but rather when longevity could potentially diverge. A linear projection of the growth shows that an unbounded longevity is expected be reached by year 2102. More advanced prediction techniques such as ARIMA or neural networks could be used, but in general these techniques are meant for short-term trends and long-term usually coincide with linear models. Table~\ref{table:shape_evo} contains per-country information regarding the projected $\xi$ evolution slope, and the corresponding boundaries of the 95\% confidence interval.

\begin{table}[h]
   \begin{tabular}{l c c c c c}
	Country & $\Delta \xi$ ($\times 10^{-4}$)  &  95\% bounds & y. div.\\
	\hline
		australia	&	4.03	&(1.22e-4,6.83e-4)&	2372	 \\
		austria	&	5.04	&(3.9e-5,9.69e-4)&	2245	\\
		belgium	&	3.10	&(2.17e-4,4.03e-4)&	2469	\\
		canada	&	6.91	&(3.93e-4,9.89e-4)&	2157	\\
		czech	&	-0.43	&(-5.47e-4,4.61e-4)&	Inf\\
		denmark	&	4.19	&(3.12e-4,5.26e-4)&	2389	\\
		finland	&	4.72	&(2.95e-4,6.49e-4)&	2321	\\
		france	&	1.55	&(5.8e-5,2.51e-4)&	2745	\\
		hungary	&	10.4	&(5.12e-4,1.56e-3)&	2128	\\
		ireland	&	11.4	&(6.68e-4,1.62e-3)&	2152	\\
		italy	&	4.34	&(3.02e-4,5.66e-4)&	2211	\\
		japan	&	-8.93	&(-1.34e-3,-4.48e-4)&	Inf	\\
		netherlands	&	4.74	&(3.63e-4,5.86e-4)&	2271	\\
		newzealand	&	-0.58	&(-6.43e-4,5.28e-4)&	Inf\\
		norway	&	7.85	&(6.74e-4,8.95e-4)&	2188	\\
		portugal	&	0.22	&(-3.86e-4,4.29e-4)&	8076\\
		slovakia	&	-3.16	&(-7.50e-4,1.18e-4)&	Inf\\
		spain	&	6.94	&(3.65e-4,1.02e-3)&	2102	\\
		sweden	&	4.04	&(3.50e-4,4.58e-4)&	2311	\\
		switzerland	&	3.59	&(2.23e-4,4.95e-4)&	2418\\
		uk	&	3.29	&(6.5e-5,5.93e-4)&	2309\\
		us	&	-3.10	&(-1.32e-5,6.96e-4)&	Inf\\
   \end{tabular}
   \caption{Shape parameter $\xi$ evolution as forecasted by a linear model. For each country, the slope and confidence intervals are detailed. Last column corresponds to the predicted year the theoretical upper bound will diverge.}
   \label{table:shape_evo}
\end{table}

\begin{figure*}[bt!]
  \begin{center}
	 \begin{tabular}{c c}
		\includegraphics[width=1\columnwidth]{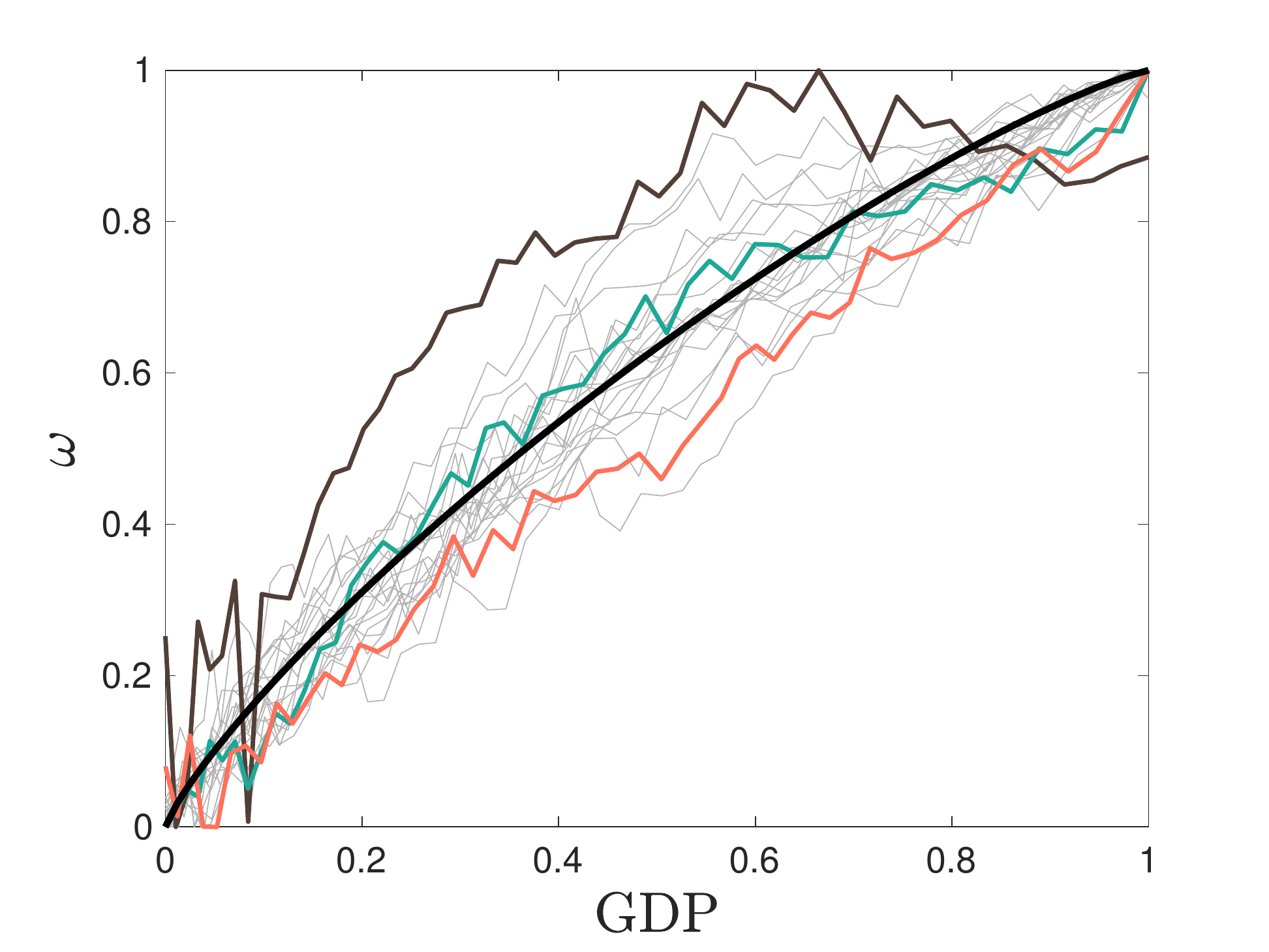} & \includegraphics[width=\columnwidth]{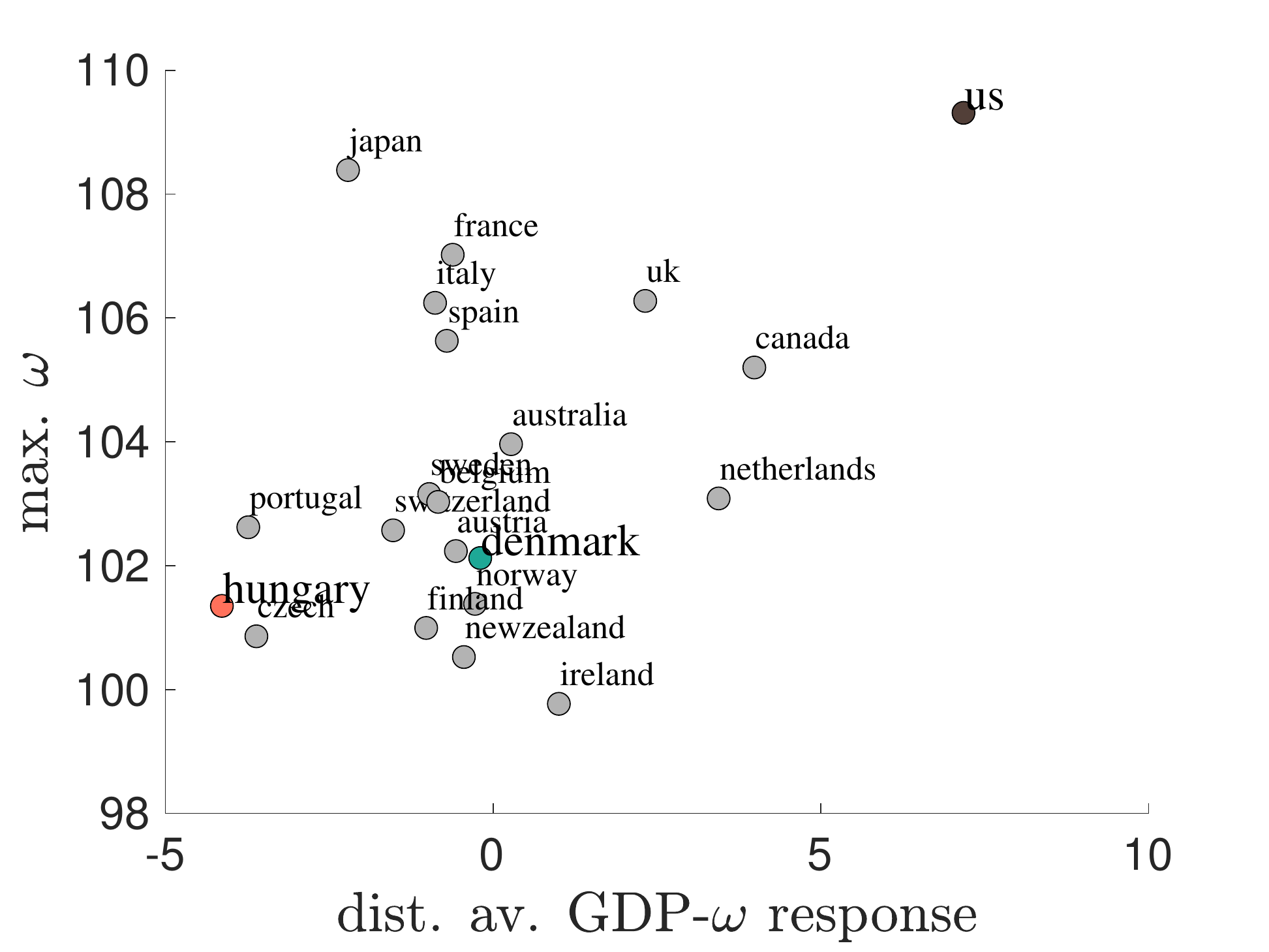}
 	 \end{tabular}
	\end{center}
	\caption{{\bf Relationship between GDP and expected maximum lifespan}. In panel A, x-axis (GDP) correspond to the accumulated GDP from 1960 to 2010 conveniently normalized in the range $[0,1]$ for each country. Y-axis corresponds to the expected maximum lifespan also conveniently normalized in the range $[0,1]$ for each country. Black solid line corresponds to the average relationship observed. The highlighted lines correspond to the two relationships with larger distances to the average, one from above and the other from below, and the most similar. Panel B shows the relationship between the distance of the observed behavior to the average and the maximum obtained expected lifespan.}
	\label{fig:dim_returns}
\end{figure*}

Figures ~\ref{fig:shape_param_linFit} and ~\ref{fig:relation_shape_maxw} may convey the (false) idea that the shape parameter $\xi$ naturally evolves independently of other factors. However, similar to life expectancy and income per capita \cite{bloom2000health}, lifespan and Gross Domestic Product (GDP) are also expected to be intertwined in some intriguing way. Differently to live expectancy, we do not expect a direct mutually reinforcing relationship since we are focusing on the tail of the live expectancy distribution; and, this contains advance age citizens that do not directly contribute to the increase of GDP. The fact that changes in the maximum lifespan --or life expectancy, for that matter-- depend ultimately on our capacity to modify ageing processes. The evolution of $\xi$ comes at a cost: the cost of decreasing our vulnerability to disease as senescence progresses. Parallel to the research lines in \cite{wilkinson1992income,wolfson1999relation,ross2000relation,stuckler2009mass,falagas2009economic}, we now relate lifespan statistics to nation-wise socio-economic indicators.  Acknowledging that GDP is just a rough proxy to how much a country invests in medical research and technology, social welfare and the adoption of healthy habits, among others, Figure~\ref{fig:dim_returns} (panel A) describes the relationship between increases in $\omega$ and increases GDP \cite{bolt2014maddison}. Note that both axis have been conveniently normalized. The general trend for all countries is a rising one, and aligned with previous findings on similar issues \cite{preston1975changing}. Interestingly, the dependence of $\omega$ on the GDP {\it does} show a saturation, which conveys the idea of diminishing returns, i.e. constant additions of the GDP yield progressively smaller increases in expected maximum lifespan.

Furthermore, we highlight in panel A three trajectories. The orange trend (lowest) corresponds to Hungary, and displays an almost linear growth of $\omega$ with GDP. The green trend (middle) corresponds to Denmark and roughly represents the average behavior (as solid black line) of the whole dataset --in the context of {\it diminishing returns}. Finally, the brown trend (top) corresponds to US, which exceptionally displays {\it negative} returns as the cumulated GDP reaches its maximum. This phenomena is typically observed when we focus on the relation of life expectation on economically-developed countries with very large GDP \cite{euromonitor2014lifeExp}.

We might expect that countries that reach larger lifespan exhibit more acute {\it diminishing returns} phenomena. To further look into this relation, Figure~\ref{fig:dim_returns}B plots the relationship between the maximum obtained lifespan on each country and the distance of the $GDP-\omega$ relationship to the average. The results evidence that there is no actual relationship between these two measures. This rises the hypothesis that the {\it diminishing returns} phenomena observed here is more related to what extent governments invest in, and citizens accept, socio-economic and technological advances to improve their health and, eventually, their lifespan.





\section{Conclusions}
As demographic data becomes more available and more reliable, our capacity to deliver better analysis can increase in multiple dimensions: deeper in the past, further in terms national or regional populations, and wider in terms of the issues relevant to demography itself (public health access, income inequality, etc.). This leap forward is possible through initiatives like the UN Data, the World Bank Open Data, or the OECD Data; but also through more specialized efforts like the Human Mortality Database or the International Database on Longevity. Here, we propose a statistical exercise that addresses the question on whether maximum human lifespan is naturally bounded --as some studies suggest--, or not. To do so, we analyse the historical trends of the maximum reported age at death for as far as 150 years, and 23 countries. This analysis is sensitive to the fact that maximum death ages are extreme values --increasingly extreme, if we consider the latest trends in age-of-death variability \cite{engelman2014lifespan}--, and as such we exploit dedicated statistical techniques, such as those that stem from Extreme Value Theory.

The resulting scenario is not that of a fixed barrier to maximum lifespan, but rather a divergent trend in longevity. The existence of an upper bound of the reversed Weibull distribution --which best represents extreme age values for an overwhelming amount of years of data-- conveys the misleading idea that there is a natural threshold to human lifespan. However, such threshold shows an increasing behavior for many countries. This suggests that our changing lifestyle and capacity to control our surrounding environment pushes forward, further every year, the mentioned barrier --which would, at least from a statistical point of view, eventually vanish. We explicitly remark that our results are the product of a statistical exercise, and as such we make no claims around the underlying biological mechanisms (genetic or otherwise) that determine human lifespan. Note, however, that these two disjoint approaches seem to converge \cite{kenyon1993c,murphy2003genes,gallego2017}.

Indeed, an evolving maximum lifespan leads to two possible interpretations: either there is no natural limit to human longevity, or such limit is not yet visible in a short- or mid-term horizon. We address this dilemma studying the changes in the shape parameter $\xi$ of the GEV distribution per calendar year. The projection of $\xi$ to the future is a helpful tool to estimate if, how, and when, longevity may diverge. If anything, lifespan limit should be studied from an evolutive and dynamical perspective.

Restraining any euphoric trust in a prospected life divergence in the near future, two facts stand out from our work. First, the stubborn singularity of the trends in ex-communist countries and US data (ironically, the most advanced industrialized nation among the studied ones), which indicate that maximum lifespan has stalled for the past decades; and, furthermore, that improvements in their economic status in the last 30 years may have rendered {\it decreased} lifespan expectations. Second, a non-linear relationship between maximum expected lifespan and GDP suggests that the fight to postpone death is increasingly costly, in the logic of diminishing returns. Our results come, admittedly, from GDP as a gross proxy of the level of development in a country. Breaking down such approximation to relevant features --public health investment, fair wealth redistribution, healthy lifestyle-- may give deeper insights on the problem but is far beyond the scope of the present work.

All of these results have, potentially, profound implications for individuals, society and the economy \cite{vaupel2010biodemography}. Divergent or not, chances are that humans will witness further increases in their maximum lifespan. As this population of centenarians and super-centenarians becomes statistically non-negligible, the challenges to sustain health and pension spending will increase sharply, raising concerns over how to provide care for increasing numbers of older people. If nothing else, the statistical examination proposed here should raise awareness among policy makers who, ultimately, must design the strategies to sustain functioning health-care systems.

\section{Methods}

\subsection{Extreme Value Theory}
The problem of modelling extreme events arises in many areas --often related to rare weather episodes, i.e. extreme floods or high wind speeds, and large financial fluctuations such as market crashes. It is in these contexts that extreme value theorem has been mostly exploited \cite{kotz2000extreme} --but also in demography \cite{gumbel1937duree,aarssen1994maximal,gbari2017extreme}. The main result from the extreme value theorem is that the maximum of a sample of i.i.d. random variables can only converge in distribution to one of 3 possible distributions, i.e. the Fr\'echet, the Gumbel, or the reversed Weibull distribution. These three are wrapped under the Generalized Extreme Value (GEV) distribution \cite{kotz2000extreme}, 

\begin{eqnarray}
	\frac{1/\sigma}t(x)^{\xi+1}e^{-t(x)}
\end{eqnarray}
where
Parameter $\mu$ controls the location of the distribution and it closely related to the first moment of the distribution. Parameter $\sigma$ determines the deviation of the distribution. And, the shape parameter $\xi$ determines to which distribution the data fit: Fr\'echet ($\xi > 0$), Gumbel ($\xi = 0$), or reversed Weibull ($\xi < 0$).


\section{Availability of source code}

Code for the analysis will be made public upon publication at the research group's webpage (http://cosin3.rdi.uoc.edu/).



\section{Availability of supporting data and materials}

The demographic datasets on mortality supporting the results of this article are publicly available in the cited repositories: HMD \cite{mortalitydb}, IDL \cite{maier2010supercentenarians}, and the Gerontology Wiki (GW) \cite{gerontology}.

Data to obtain the seasonal birth and death distribution can be queried on UNdata (http://data.un.org/) webpage, under the label ``Death by month of death''. 

Regarding economic data, historical per-country GDP data is available from the Maddison Project \cite{bolt2014maddison}.

Additional materials: the Supplementary Information file (pdf) contains two figures, Figure S1 and Figure S2, along with some explanatory text.

\section{Declarations}

Not applicable.

\subsection{Consent for publication}

Not applicable.

\subsection{Competing Interests}

The authors declare that they have no competing interests.



\subsection{Author's Contributions}

All authors contributed equally to the paper.

\end{document}


\title{Supplementary Information}

\author{Albert Sol\'e-Ribalta, Javier Borge-Holthoefer}

\maketitle



\subsection*{I. Average maximum lifespan as a function of population size}
This section analyses the dependance of our results with respect to the sampling procedure and the underlying population size. In other words, we inspect here to what extent maximum lifespan is affected by country population size. 

Figure S1 plots the expected maximum lifespan (as per fitted distribution) against population size. Each country is represented as a different colour, and results correspond to the period 1960-2014. The left-most country (dark red) is New Zealand; on the right, grey dots correspond to Japan, and black ones (right-most) to U.S. For every country, the lowest $\omega$ value corresponds to $t=1960$. In general, the highest $\omega$ corresponds to $t=2014$ (with the exception of U.S.). 

The main conclusion from this figure is the lack of an obvious relationship between maximum lifespan and population. Taken individually, all countries (except U.S.) have witnessed dramatic increases in their maximum lifespan expectations, {\em regardless} of their population growth patterns. For instance, New Zealand (which doubled its population in the 1960-2014 period), Hungary (with a receding population in the same period), and Japan (with a very stable population in the last 40 years) show similar trends regarding maximum lifespan. In the opposite  trend we find U.S.: doubling its population in the period 1960-2014, its maximum expected lifespan has overall increased --though very moderately--, appears to have reached a plateau (and even, at times, to decrease).

\begin{figure*}[h!]
  \begin{center}
    \includegraphics[width=\columnwidth]{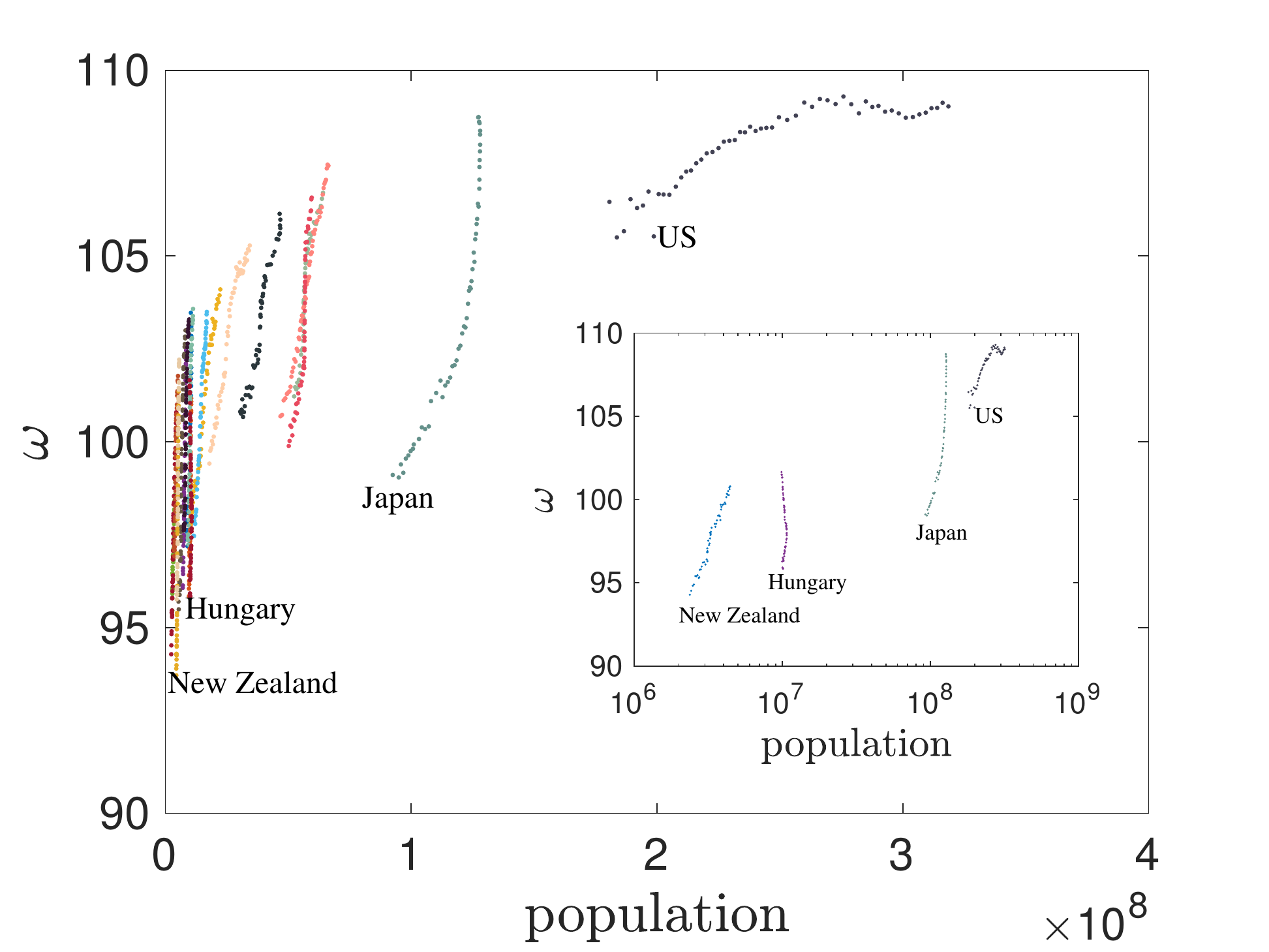}        
  \end{center}
	\caption{Expected maximum lifespan (as per fitted distribution) against population size. Each country is represented as a different colour, and results shown for the period 1960-2014. The left-most country (dark red)  is New Zealand; on the right, grey dots correspond to Japan, and black ones (right-most) to U.S. For every country, the lowest $\omega$ value corresponds to $t=1960$. In general, the highest $\omega$ corresponds to $t=2010$ (with the exception of U.S.). To facilitate visual inspection, the inset shows the same data, for the mentioned countries, in logarithmic scale.}
	\label{fig:dim_returns}
\end{figure*}

\clearpage

\subsection*{II. Life expectancy at ages 0-90}
Figure~S2 show the evolution of life expectancy at different ages, for four countries that have also been analyzed in the main text under the light of GEV. Clearly, these plots fail to grasp the boldly increasing behaviour of average maximum lifespan --which is evident in Figure~1 of the main text. This indicates that, although valuable to anticipate general trends for the bulk of the population, these tools are not meant to capture the behaviour of extreme scores.

\begin{figure*}[h]
  \begin{center}
	 \begin{tabular}{c}
		\includegraphics[width=1\columnwidth]{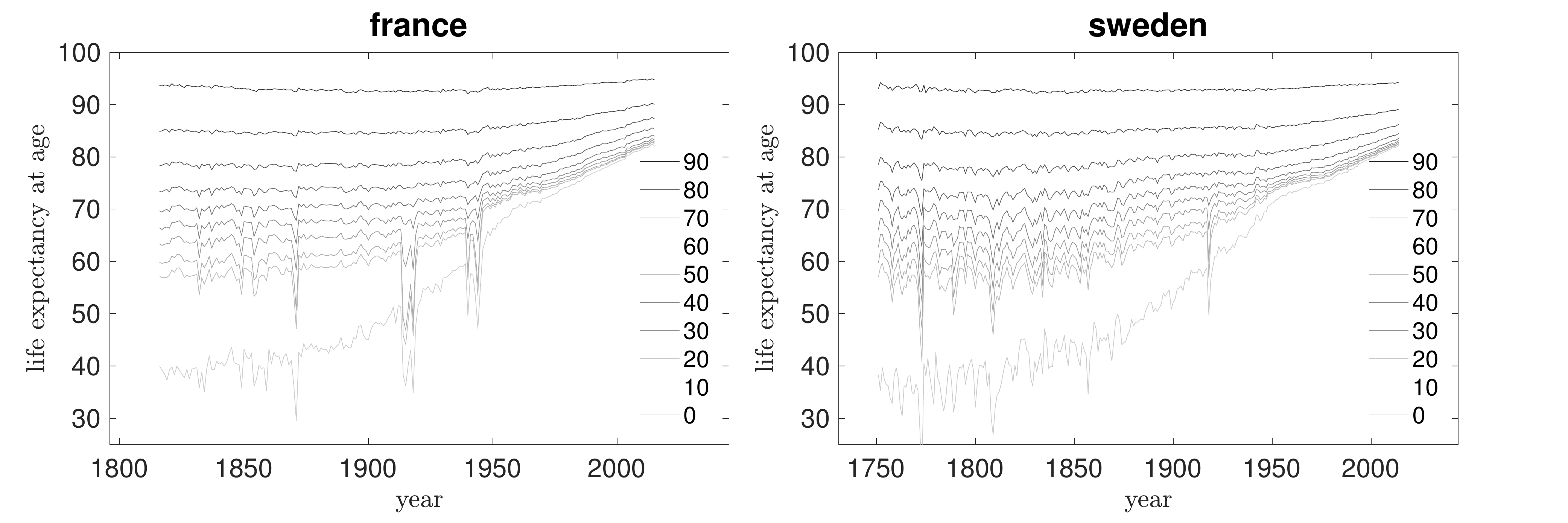}\\
		 \includegraphics[width=\columnwidth]{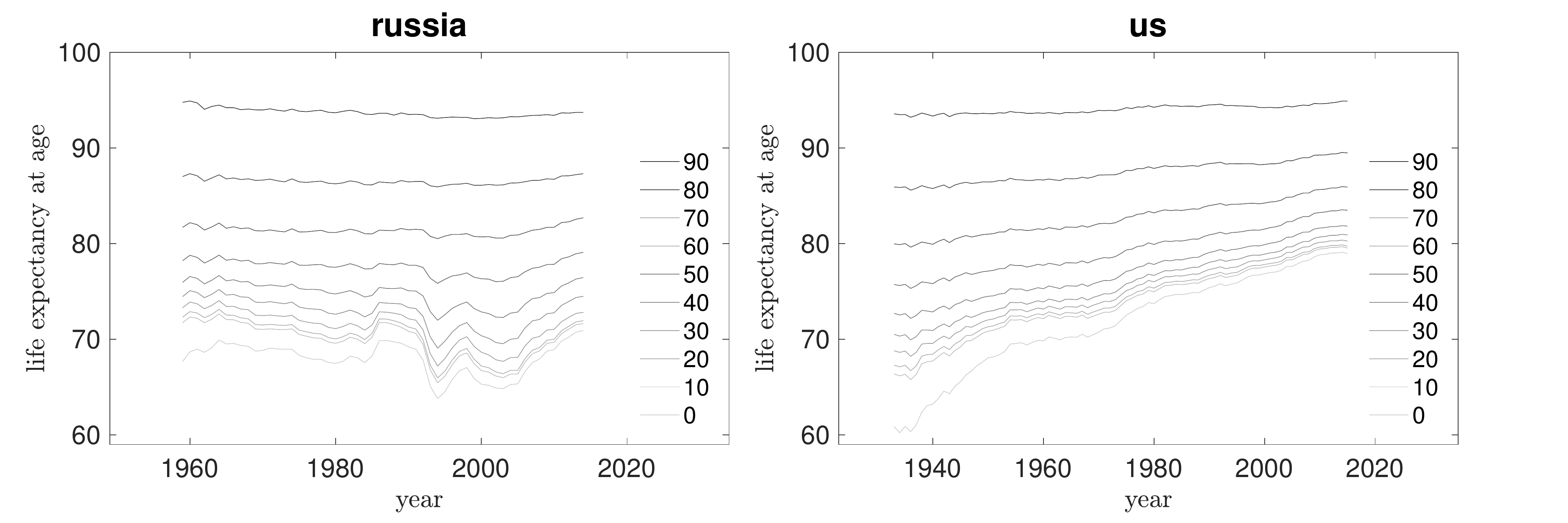}
 	 \end{tabular}
	\end{center}
	\caption{Evolution of the life expectancy at age for four of the countries analysed in the research work. The data is extracted from The Human Mortality Database (http://www.mortality.org)}
	\label{fig:dim_returns}
\end{figure*}
